\documentclass[aps,prl,reprint,footinbib,longbibliography]{revtex4-2}

\usepackage{amsmath,amsthm,amsfonts,amssymb,bm,bbold}
\usepackage{mdframed,graphicx,array,MnSymbol,caption,subcaption,float,placeins}
\usepackage[usenames,dvipsnames]{xcolor}
\definecolor{OceanBlue}{rgb}{0,0.35,0.7} 
\usepackage{hyperref}
\hypersetup{colorlinks=true,allcolors=OceanBlue}

\newtheorem{lemma}{Lemma}[section] 

\theoremstyle{definition}

\theoremstyle{remark}
\newmdenv[
  topline=false,
  bottomline=false,
  rightline=false,
  skipabove=\topsep,
  skipbelow=\topsep,
  leftmargin=-5pt,
  rightmargin=-10pt,
  innertopmargin=0pt,
  innerbottommargin=0pt
]{leftrule} 

\newcommand{\pars}[1]{\left(#1\right)} 
\newcommand{\bracs}[1]{\left[#1\right]} 

\newcommand{\mr}{\mathrm} 
\newcommand{\mb}{\mathbf} 
\newcommand {\cl}{\mathcal} 
\newcommand {\tsf} [1]{\textsf{#1}} 
\newcommand{\vectsym}{\boldsymbol}  
\DeclareMathOperator{\Tr}{Tr\,}            
\DeclareMathOperator{\tr}{tr\,}            

\newcommand{\diff}{\mathop{}\!\mathrm{d}}
\DeclareMathOperator{\trans}{\textsf{T}} 
\DeclareMathOperator{\sech}{\mathrm{sech}} 
\DeclareMathOperator{\csch}{\mathrm{csch}} 

\newcommand{\balpha}{\vectsym{\alpha}}  
\newcommand{\bbeta}{\vectsym{\beta}} 

\newcommand {\ket}[1] {\left|{#1}\right\rangle}
\newcommand {\kets} [1] {\left|#1\right\rangle_{S}}
\newcommand {\kete} [1] {\left|#1\right\rangle_{E}}
\newcommand {\keta} [1] {\left|#1\right\rangle_{A}}

\newcommand {\bra}[1] {\langle{#1}|}
\newcommand{\braket}[2]{\langle{#1}|{#2}\rangle}

\newcommand{\varket}[1] {\left|{#1}\right\rrangle}
\newcommand{\varbra}[1] {\left\llangle #1\right|}
\newcommand{\varbraket}[2] {\left\llangle #1| #2 \right\rrangle}

\newcommand{\abs}[1]{\left | #1 \right |}   
\newcommand{\mean}[1]{\left\langle #1 \right\rangle}  

\begin{document}
\title{Optimal gain sensing of quantum-limited phase-insensitive amplifiers}
\author{Ranjith Nair$^{1,2}$}
\email{ranjith.nair@ntu.edu.sg}
\author{Guo Yao Tham$^{1}$}
\email{ tham0157@e.ntu.edu.sg}
\author{Mile Gu$^{1,2,3}$}
\email{gumile@ntu.edu.sg}
\affiliation{$^1$Nanyang Quantum Hub, School of Physical and Mathematical Sciences,  \\
 Nanyang Technological University, 21 Nanyang Link, Singapore 639673 \\
$^2$Complexity Institute, Nanyang Technological University, 61 Nanyang Drive, Singapore 637460 \\
$^3$Centre for Quantum Technologies, National University of Singapore, 3 Science Drive 2, Singapore 117543}

\date{\today}
\begin{abstract} Phase-insensitive optical amplifiers uniformly amplify each quadrature of an input field and are  of both fundamental and technological importance.  We find the quantum limit on the precision of estimating the gain of a quantum-limited phase-insensitive optical amplifier using a multimode probe  that may also be  entangled with an ancilla system.   In stark contrast to the sensing of loss parameters, the  average photon number $N$ and number of input modes $M$ of the probe are found to be equivalent and interchangeable resources for optimal  gain sensing. All pure-state probes whose reduced state on the input modes to the amplifier is diagonal in the multimode number basis are proven to be quantum-optimal under the same gain-independent measurement.  We compare the best precision achievable using classical probes to the performance of an explicit photon-counting-based estimator on quantum probes and show that an advantage exists even for single-photon probes and inefficient photodetection.  A closed-form expression for the energy-constrained Bures distance between two product amplifier channels is also derived.
\end{abstract}
\maketitle


Phase-insensitive  amplifiers coherently and uniformly amplify every quadrature  amplitude of an input electromagnetic field. The prototypical example of such an amplifier is a laser gain medium with population inversion between the active levels. Besides being a key component of lasers, phase-insensitive amplifiers (e.g., erbium-doped fiber amplifiers) are widely deployed in today's optical communication networks for restoring signal amplitudes and to offset detection noise \cite{RSS09optical}. Many physical mechanisms leading to phase-insensitive amplification are known in diverse platforms (see, e.g., Refs.~\cite{CDG+10,CCJ+12,LVH+14,CHN+20}), but they are all constrained by the unitarity of  quantum dynamics to add a gain-dependent excess noise \cite{HM62,Cav82,CDG+10} that is minimized when the effective population in the active levels of the gain medium is completely inverted \cite{CCJ+12,Aga12quantum}. 

Such minimum-noise phase-insensitive amplifiers -- hereafter called \emph{quantum-limited amplifiers (QLAs)} -- are also of fundamental importance in continuous-variable quantum information. This is because the quantum channels defined by QLAs,  together with  pure-loss  channels, are building blocks for constructing all other phase-covariant Gaussian channels  by concatenation \cite{CGH06,Ser17qcv}. Due to the ubiquity of loss channels in nature, there is a vast literature on their sensing (see, e.g., \cite{Nai18loss,PBG+18,BAB+18,PVS+20} and references therein). In contrast, previous work on sensing gain of a QLA is limited to the context of detecting Unruh-Hawking radiation using single-mode probes \cite{AAF10} or assumes access to the internal degrees of freedom of the amplifier \cite{GP09}. 

In this Letter, we fill this gap by optimizing the gain sensing precision over all multimode ancilla-entangled probes and all joint quantum measurements, constraining only the energy and number of input modes of the probe.  We also propose concrete probes,  measurements and estimators enabling laboratory demonstration of a quantum advantage using present-day technology limited by nonunity-efficiency photodetection. Beyond gain sensing itself, owing to the above-mentioned concatenation theorem, our results  combined with those for pure-loss channels \cite{Nai18loss} are expected to yield fundamental performance limits for a vast suite of detection and estimation problems involving Gaussian channels with excess noise -- see, e.g., Refs. \cite{TEG+08,NG20,BCT+21,GMT+20,Pir11,OLP+21,MI11,SWA+18,SSW20arxiv,WDA20,JDC22,ZP20,HBP21,Zhu21,BGD+17,TBG+21}.

\begin{figure}[btp]
\centering\includegraphics[trim=55mm 82mm 35mm 78mm, clip=true,width=\columnwidth]{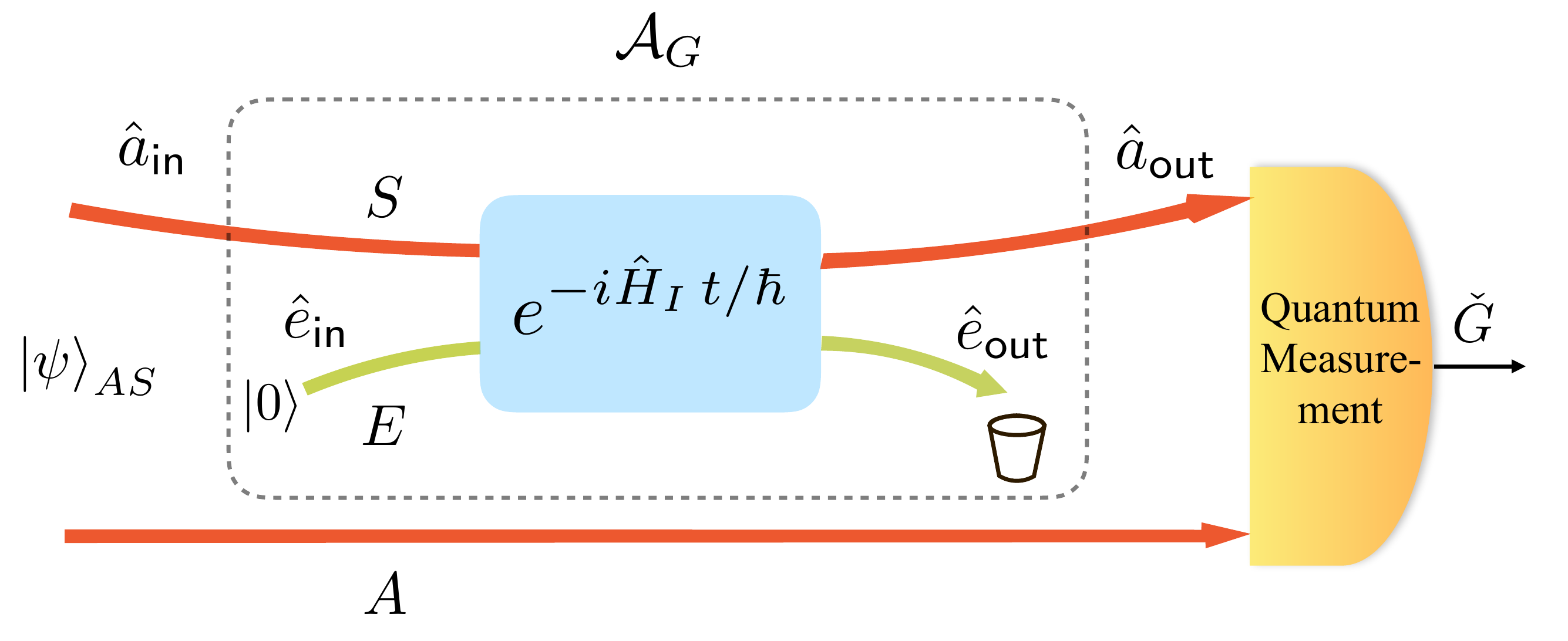}
\caption{A general ancilla-assisted parallel strategy for sensing the gain $G$ of a QLA $\cl{A}_G$ (dashed box): Each of $M$ signal ($S$) modes (one of which is shown) of a probe $\ket{\psi}_{AS}$ possibly entangled with an ancilla system $A$ is subject to a two-mode squeezing interaction $\hat{H}_I = i\hbar\kappa\pars{\hat{a} \hat{e} -\hat{a}^\dag \hat{e}^\dag}$ between the $S$ mode (annihilation operator $\hat{a}$) and an environment ($E$) mode (annihilation operator $\hat{e}$) initially in the vacuum state. An estimate $\check{G}$ of $G = \cosh^2\kappa t$ is obtained using the optimal joint measurement on the output of $AS$.}\label{fig:ampgainsensing}
\end{figure}

\indent \emph{Quantum-limited amplifiers--} A canonical realization of a QLA involves an optical parametric amplifier (or \emph{paramp}) effecting a two-mode squeezing interaction  between the amplified or  \emph{signal} ($S$) mode (annihilation operator $\hat{a}$) and an \emph{environment} mode  ($E$) (annihilation operator $\hat{e}$), after which the $E$ mode is discarded \cite{Aga12quantum,CCJ+12,Ser17qcv} (Fig.~\ref{fig:ampgainsensing}, dashed box). In the interaction picture, the paramp Hamiltonian $\hat{H}_I = i\hbar \kappa \pars{\hat{a} \hat{e} -\hat{a}^\dag \hat{e}^\dag }$, where $\kappa$ is an effective coupling strength. Quantum-limited operation obtains when the environment is initially in the vacuum state.  Evolution for a time $t$ results in the Bogoliubov transformations
$\hat{a}_\tsf{out} = \sqrt{G} \,\hat{a}_\tsf{in} - \sqrt{G-1}\,\hat{e}^\dag_\tsf{in};
\hat{e}_\tsf{out}  = \sqrt{G} \,\hat{e}_\tsf{in} - \sqrt{G -1}\,\hat{a}^\dag_\tsf{in},
$
where $\hat{a}_\tsf{in} = \hat{a}(0), \hat{e}_\tsf{in} =\hat{e}(0)$ are the input (time zero) and $\hat{a}_\tsf{out} = \hat{a}(t), \hat{e}_\tsf{out} =\hat{e}(t)$ are the output (time $t$) annihilation operators and $\sqrt{G} = \cosh \kappa t \equiv \cosh\tau$.  The average output energy $\mean{\hat{a}^\dag_\tsf{out} \hat{a}_\tsf{out}} = G\mean{\hat{a}^\dag_\tsf{in} \hat{a}_\tsf{in}}+ (G-1)$, where the last term represents the added noise of a QLA of gain $G \geq 1$ \cite{Cav82}.  
The state transformation (quantum channel) on the signal mode corresponding to a QLA of gain $G$ is denoted $\cl{A}_G$.

\indent \emph{Gain sensing setup and background ---} Figure~\ref{fig:ampgainsensing} also shows a general ancilla-assisted parallel estimation strategy for gain sensing. A pure state  $\ket{\psi}_{AS}$ (called the \emph{probe}) of $M$ signal modes entangled with an arbitrary ancilla system $A$ is prepared.  Each of the signal modes passes through the QLA, following which the joint $AS$ system is measured using an optimal (possibly probe-dependent) measurement for estimating $G$.  
 The probe has the general form 
\begin{align} \label{probe}
\ket{\psi}_{AS} = \sum_{\mb{n} \geq \mb{0}} \sqrt{p}_{\mb{n}}\ket{\chi_{\mb n}}_A \ket{\mb n}_S,
\end{align}
where $\ket{\mb n}_S = \ket{n_1}_{S_1}\ket{n_2}_{S_2}\cdots \ket{n_M}_{S_M}$ is an $M$-mode number state of $S$, $\{\ket{\chi_{\mb n}}_A\}$ are normalized (not necessarily orthogonal) states of $A$, and $\left\{p_{\mb n} \geq 0 \right\}$ is the probability distribution of $\mb{n}$. 
The number $M$ of available signal modes depends on operational constraints such as measurement time and bandwidth, and will turn out to be fundamental in determining the sensing precision.
Additionally, we impose the standard  constraint on the average photon number in the signal modes: $\bra{\psi}  \hat{I}_A \otimes \pars{\sum_{m=1}^M \hat{N}_m} \ket{\psi} = N$, where $\hat{N}_m = \hat{a}_m^\dag \hat{a}_m$ is the number operator of the $m$-th signal mode and $\hat{I}_A$ is the identity on the ancilla system. This constraint can be simplified as 
$\sum_{n=0}^{\infty} n\,p_n = N,\;\;{\rm where}\;\; p_n = \sum_{\mb{n}\, :\, n_1 +\ldots + n_M = n} p_{\mb n}
$
is the probability mass function of the \emph{total} photon number in the signal modes. A mixed-state probe can be purified using an additional ancilla with the resulting purification being again of the form ~\eqref{probe} with the same $N$ and $M$. Thus, optimization over probes of the form of Eq.~\eqref{probe} suffices.

We are interested in comparing the performance of the optimal quantum probes of the form of Eq.~\eqref{probe} to the best performance achievable using \emph{classical} probes under the same resource constraints, i.e., probes that consist of mixtures of $M$-mode coherent states, possibly correlated with an arbitrary number $M'$ of ancilla modes. Such probes can be prepared using laser sources, and have the form
\begin{align}\label{classicalprobe}
\rho_{AS} = \iint \diff^{2M'}\balpha\diff^{2M} \bbeta \,P(\balpha,\bbeta) \ket{\balpha}\bra{\balpha}_A\otimes\ket{\bbeta} \bra{\bbeta}_S,
\end{align}
where $\balpha = \pars{\alpha^{(1)}, \ldots, \alpha^{(M')}} \in \mathbb{C}^{M'}$ and $\bbeta = \pars{\beta^{(1)}, \ldots, \beta^{(M)}}\in \mathbb{C}^{M}$ index $M'$- and $M$-mode coherent states of $A$ and $S$ respectively,  and $P\pars{\balpha,\bbeta} \geqslant 0$ is a probability distribution. The signal energy constraint  takes the form  
$
\int_{\mathbb{C}^{M'}} \diff^{2M'} \balpha \int_{\mathbb{C}^M} \diff^{2M} \bbeta\, P(\balpha,\bbeta) \pars{\sum_{m=1}^M \abs{\bbeta^{(m)}}^2}  = {N}.
$
  

Given a probe $\ket{\psi}_{AS}$, we have the output state  $\rho_G :=  {\mr{id}_A \otimes \cl{A}_G^{\otimes M}}\pars{\ket{\psi}\bra{\psi}_{AS}}$ ($\rho_G = \mr{id}_A \otimes \cl{A}_G^{\otimes M}\pars{\rho_{AS}}$ for a classical probe \eqref{classicalprobe}), where $\mr{id}_A$ is the identity channel on $A$.
Estimation of $G$ from the state family $\{\rho_{G}\}$ is subject to the \emph{quantum Cram\'er-Rao bound} (QCRB) \cite{Hel76,*Hol11,PBG+18,PVS+20}, a brief description of which  follows.  A measurement on the $AS$ system is described by a collection of positive operators $\left\{\hat{\Pi}_y\right\}_{\cl{Y}}$ indexed by the measurement result $y \in \cl{Y}$ and summing to the identity. The probability distribution of the result   $P(y;G) = \Tr \rho_G \hat{\Pi}_y$, and an estimator $\check{G}(y)$ based on this measurement is called \emph{unbiased} for $G$ if  $\int_{\cl{Y}} \diff  y\, \check{G}(y) P(y;G) = G$ for all $G$ in the interval of interest. The (classical) \emph{Cram\'er-Rao bound} (CRB)  bounds the mean squared error (MSE) $\mathbb{E}\bracs{\check{G} - G}^2$ of any unbiased estimator as $
\mathbb{E}\bracs{\check{G} - G}^2 \geq  1/{\cl{J}_G \bracs{Y}}$, where
$\cl{J}_G \bracs{Y} := \mathbb{E} \bracs{{\partial_{G} \ln P(Y;G)}}^2 = - \mathbb{E} \bracs{{\partial_{G}^2 \ln P(Y;G)}},
$
 is the (classical) \emph{Fisher information} (FI) on $G$ of the measurement $Y$ \cite{Kay93ssp1}.
Different measurements $\left\{\hat{\Pi}_y\right\}_{\cl{Y}}$ result in different CRBs. 

On the other hand, there exists a Hermitian operator $\hat{L}_G$ called the symmetric logarithmic derivative (SLD)  satisfying
 $\partial_G \rho_{G} \equiv \partial \rho_{G}/ \partial G = \pars{\rho_{G} \hat{L}_{G} + \hat{L}_{G} \rho_{G}}/2.$
 The \emph{quantum Fisher information}  (QFI) is defined as 
  $\cl{K}_G = \Tr \rho_G \hat{L}_{G}^2,$ and 
the QCRB 
$
\mathbb{E}\bracs{\check{G} - G}^2 \geq \cl{K}_G^{-1}
$
minimizes the right-hand side of the CRB over all unbiased measurements and defines the quantum-optimal sensing performance. The QFI $\cl{K}_\theta$ on $\theta$ of an arbitrary state family $\{\rho_{\theta}\}$ is  related to the fidelity  $F\pars{\rho_\theta, \rho_{\theta'}} = \Tr \sqrt{\sqrt{\rho_{\theta}} \rho_{\theta'} \sqrt{\rho_{\theta}}}$ between the states of the family via \cite{Hay06,*BC94}
\begin{align} \label{QFIfromfidelity}
\cl{K}_{\theta} =  -4\, \partial^2_{\theta'} F\pars{\rho_\theta,\rho_{\theta'}}\vert_{\theta'=\theta}. 
\end{align} It is expedient for us to work with the QFI on the parameter $\tau = \kappa t =\mr{arccosh} \sqrt{G}$. Since the SLDs with respect to $\tau$ and $G$ satisfy $\hat{L}_G = \frac{\partial \tau}{\partial G}\, \hat{L}_\tau$,  $\cl{K}_G = \pars{\frac{\partial \tau}{\partial G}}^2 \cl{K}_\tau$ so that maximizing either QFI suffices.

\indent \emph{Optimal gain sensing ---}
We first obtain an upper bound  $\widetilde{\cl{K}}_\tau \geq {\cl{K}}_\tau$ on the QFI  in  the hypothetical situation where the  output  of $ASE$ is available for measurement. Given a probe $\ket{\psi}_{AS}$, we then hold the state family $\left\{ \Psi_{\tau} = \ket{\psi_\tau} \bra{\psi_\tau}\right\}$ defined by 
$
\ket{\psi_{\tau}}_{ASE} = \hat{I}_A \otimes  \pars{\otimes_{m=1}^{M} \hat{U}_m(\tau)} \ket{\psi}_{AS}\ket{\mb{0}}_E,
$
where  $\hat{U}_m(\tau)=\exp\pars{-i\hat{H}_I t/ \hbar}$ is the paramp unitary  acting on the $m$-th signal and environment mode pair parametrized by $\tau = \kappa t$. 
The QFI $\widetilde{\cl{K}}_{\tau}$ of $\left\{\Psi_{\tau}\right\}$  upper bounds $\cl{K}_\tau$ due to the monotonicity of the QFI with respect to partial trace over $E$ \cite{Hay06}. 

We can show (see Appendix A) that the paramp takes the input $\ket{n}_S\ket{0}_E$ to
$\hat{U}(\tau) \kets{n}\kete{0}= 
 \sech^{(n+1)} \tau\sum_{a=0}^\infty \sqrt{{n +a \choose a}} \tanh^a \tau \kets{n+a}\kete{a}.$
Thus, the paramp coherently adds a random number $a$ of photons to both $S$ and 
$E$ according to the negative binomial distribution $\mr{NB}(n+1, \sech^2 \tau)$ \cite{RS15probstats}. For any probe \eqref{probe}, we have
\begin{align} \label{ASEstate2}
\ket{\psi_\tau}_{ASE} &=  \sum_{\mb{a} \geq \mb{0}}  \varket{\psi_{\mb{a};\tau}}_{AS} \ket{\mb{a}}_E,
\end{align}
where
$\varket{\psi_{\mb{a};\tau}}_{AS} = \sum_{\mb{n} \geq\mb{0}}  \sqrt{p_\mb{n} A_\tau(\mb{n},\mb{a})} \ket{\chi_{\mb{n}}}_A\ket{\mb{n} +\mb{a}}_S
$
are non-normalized states of $AS$ and $A_\tau(\mb{n},\mb{a}) = \prod_{m=1}^M {n_m + a_m \choose a_m} \sech^{2(n_m+1)}\tau \tanh ^{2 a_m}\tau$ is a product of negative binomial probabilities. For $\ket{\psi_{\tau'}}_{ASE}$  the  output state on $ASE$ obtained by passing $\ket{\psi}_{AS}$  through a QLA of gain $G' = \cosh^2 \tau'$, the fidelity between the  outputs can be shown after some computation (see Appendix B) to be  
\begin{align} \label{ASEoverlap}
F\pars{\Psi_\tau, \Psi_{\tau'}} =\braket{{\psi_\tau}}{{\psi_{\tau'}}} &=  \sum_{n=0}^\infty p_n \nu^{n+M}, 
\end{align}
where $\nu =\sech\,(\tau'-\tau) = \pars{\sqrt{GG'} - \sqrt{(G-1)(G'-1)}}^{-1}\in (0,1]$. Using Eq.~\eqref{QFIfromfidelity}, we obtain the sought upper bounds
$\cl{\widetilde{K}}_\tau = 4(N+M)$ and 
$ \widetilde{\cl{K}}_G  = \frac{N+M}{G(G-1)}$
on the true QFI with respect to $\tau$ and $G$.

Returning to the original problem in which only $\rho_\tau = \Tr_E \Psi_\tau$  is accessible, we have from Eq.~\eqref{ASEstate2} that
$
\rho_\tau =  \sum_{\mb{a}\geq \mb{0}} \varket{\psi_{\mb{a};\tau}} \varbra{\psi_{\mb{a};\tau}}_{AS}.
$
 For given $\left\{p_{\mb{n}}\right\}$ in Eq.~\eqref{probe}, consider probes  for which  $\{\ket{\chi_{\mb n}}_A\}$ is an orthonormal set. Such probes, called \emph{number-diagonal signal (NDS)} probes, are known to be optimal probes for diverse sensing problems \cite{NY11,SWA+18,Nai18loss}. Orthonormality of  the $\{\ket{\chi_{\mb n}}_A\}$ implies that $\varbraket{\psi_{\mb{a};\tau}}{\psi_{\mb{a'};\tau'}} = \varbraket{\psi_{\mb{a};\tau}}{\psi_{\mb{a};\tau'}}  \delta_{\mb{a},\mb{a'}}$, so the output fidelity 
$
F\pars{\rho_\tau,\rho_{\tau'}} = \sum_{\mb{a}\geq\mb{0}} \varbraket{\psi_{\mb{a};\tau}}{\psi_{\mb{a};\tau'}} =  F\pars{\Psi_\tau, \Psi_{\tau'}}$ of Eq.~\eqref{ASEoverlap}. Thus, the QFIs  on $\tau$ and $G$
 \begin{align} \label{ndsqfi}
\cl{{K}}_\tau = 4(N+M);\;\;
 {\cl{K}}_G  = \frac{N+M}{G(G-1)}
 \end{align}
 of NDS probes saturate the upper bounds calculated above.
 
This result exhibits several remarkable features. First, {any} NDS probe with the given $N$ and $M$ is  quantum-optimal  regardless of its exact  signal photon number distribution $\left\{p_{\mb{n}}\right\}$. This generalizes the single-mode Fock-state optimality result \cite{AAF10} not just to multimode Fock states but to the infinite class of ancilla-entangled multimode NDS probes including the workhorse of optical quantum information -- the two-mode squeezed vacuum (TMSV) state. Second,  gain sensing performance explicitly depends on the number $M$ of signal modes.  This contrasts sharply with loss sensing, for which the optimal QFI is $M$-independent \cite{Nai18loss}. Physically, this difference stems from the gain-dependent quantum noise introduced by a QLA that makes the output states of two QLAs with distinct gains distinguishable even for a vacuum input. Increasing the number of signal modes further improves their distinguishability. In contrast, vacuum probes of any $M$ are invariant states of loss channels and are therefore useless for sensing them. Finally, the roles of $N$ and $M$ in Eq.~\eqref{ndsqfi} are seen to be equivalent so that one resource can be exchanged for the other, providing additional flexibility in the choice of optimal probes.

\begin{figure}[t]
\centering
  \includegraphics[trim=12.5mm 60mm 22mm 72mm, clip=true, scale=0.44]{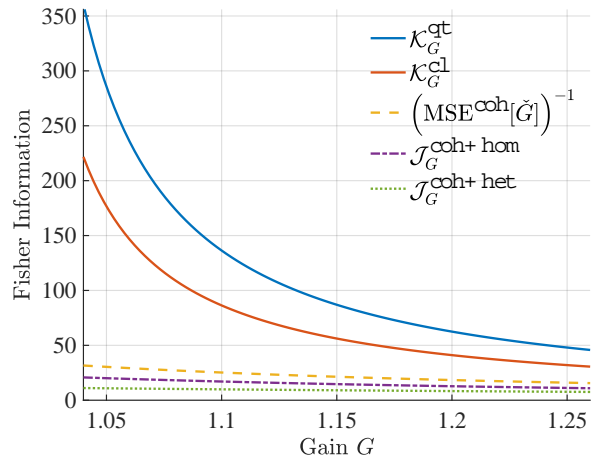}
  \caption{The optimal quantum (blue) (Eq.~\eqref{ndsqfi}) and classical QFI (red) (Eq.~\eqref{csQFI}) for  $N=6$ and $M=9$. Also shown are the FI of homodyne  (purple dashed-dotted), heterodyne detection (green dotted), and the inverse MSE of the photodetection-based estimator (Eq.~\eqref{Gcheck}) (yellow dashed) for a coherent-state probe.}
  \label{fig:totalfi}
\end{figure}

For  an $M$-mode signal-only coherent-state probe $\ket{\sqrt{N_1}}_{S_1}\cdots\ket{\sqrt{N_M}}_{S_M}$ with $\sum_{m=1}^M N_m = N$, the  output state $\rho_\tau$ is a product of single-mode Gaussian states. The QFI on $G$ follows from the results of  \cite{PJT+13} after some algebra:
\begin{equation}
\begin{aligned} \label{csQFI}
\cl{K}^{\tsf{coh}}_G &= {\frac{N}{G(2G-1)}  + \frac{M}{G(G-1)}}. 
\end{aligned}
\end{equation} 
 The convexity of QFI in the state \cite{Fuj01} and the linear dependence on $N$ of the first term in the above expressions  imply that no classical probe [Eq.~\eqref{classicalprobe}] with $M$ signal modes can beat the QFI of Eq.~\eqref{csQFI}.
Both  Eqs.~\eqref{ndsqfi} and \eqref{csQFI} contain a term proportional to $N$ (the \emph{photon contribution}) and another proportional to $M$ (the \emph{modal contribution}). The modal contribution in the optimal quantum and classical QFI is identical, but the quantum-optimal photon contribution is at least twice the classical photon contribution and far exceeds it in the $G \sim 1$ regime (see Fig.~\ref{fig:totalfi}).

  \emph{Performance of standard measurements---} 
Suppose that an arbitrary NDS probe \eqref{probe} is input to a QLA of unknown gain and that we measure the basis $\left\{\keta{\chi_{\mb n}}\right\}$  and also the photon number in each of the $M$ output signal modes. Denote the measurement result $(\mb{X},\mb{Y})$, where  $\mb{X} = (X_1, \ldots, X_M)$ if $\ket{\chi_{\mb{X}}}_A$ is the measurement result on $A$ and $\mb{Y} = (Y_1,\ldots, Y_M)$ if $Y_m$ photons are observed in the $m$-th output signal mode. 
We can then show (Appendix C) that the FI
$\cl{J}_\tau \bracs{\mb{X},\mb{Y}}  = 4(N+M)$ 
for any NDS probe, so that this measurement achieves the quantum-optimal QFI \eqref{ndsqfi}. 

While this implies that the maximum likelihood estimator based on $(\mb{X},\mb{Y})$ achieves the quantum limit for a large number of copies \cite{RS15probstats,Kay93ssp1},  a quantum-optimal estimator may not exist for a finite sample \cite{Kay93ssp1}. For a multimode number-state probe $\otimes_{m=1}^M \ket{n_m}_{S_m}$ with $\sum_{m=1}^M n_m = N$, consider the estimator 
\begin{align} \label{Gcheck}
\check{G} := \pars{Y+ M}/\pars{N+M},
\end{align}
where $Y = \sum_{m=1}^M Y_m$ is the total photon number measured in the signal modes. Using the fact that $Y - N \thicksim \mr{NB}(N+M, \sech^2 \tau)$, we can show that $\check{G}$ is unbiased and that $\mr{Var}[\check{G}] = \frac{G(G-1)}{N+M}$ so the QCRB \eqref{ndsqfi} is achieved even on a finite sample for any multimode number-state probe.

On the other hand, a $G$-independent measurement that achieves the coherent-state QFI \eqref{csQFI} is unknown. The estimator $\check{G}$ above remains unbiased but has the suboptimal variance $\mathrm{MSE}^{\textsf{coh}}[\check{G}] = \frac{G(G-1)}{N+M} + \frac{G^2 N}{(N+M)^2}$ (Appendix D.~III). Homodyne and heterodyne  detection in each output mode have the respective (suboptimal) FIs $\cl{J}^{\tsf{coh+hom}}_G = \frac{N}{G(2G-1)} + \frac{2M}{(2G-1)^2}$ and
$\cl{J}^{\tsf{coh+het}}_G = \frac{N/2 + M}{G^2}.$
These Fisher information quantities are compared in Fig.~\ref{fig:totalfi}.

\begin{figure}[t]
\centering
  \includegraphics[trim=54mm 98mm 106mm 92mm, clip=true, scale=0.4]{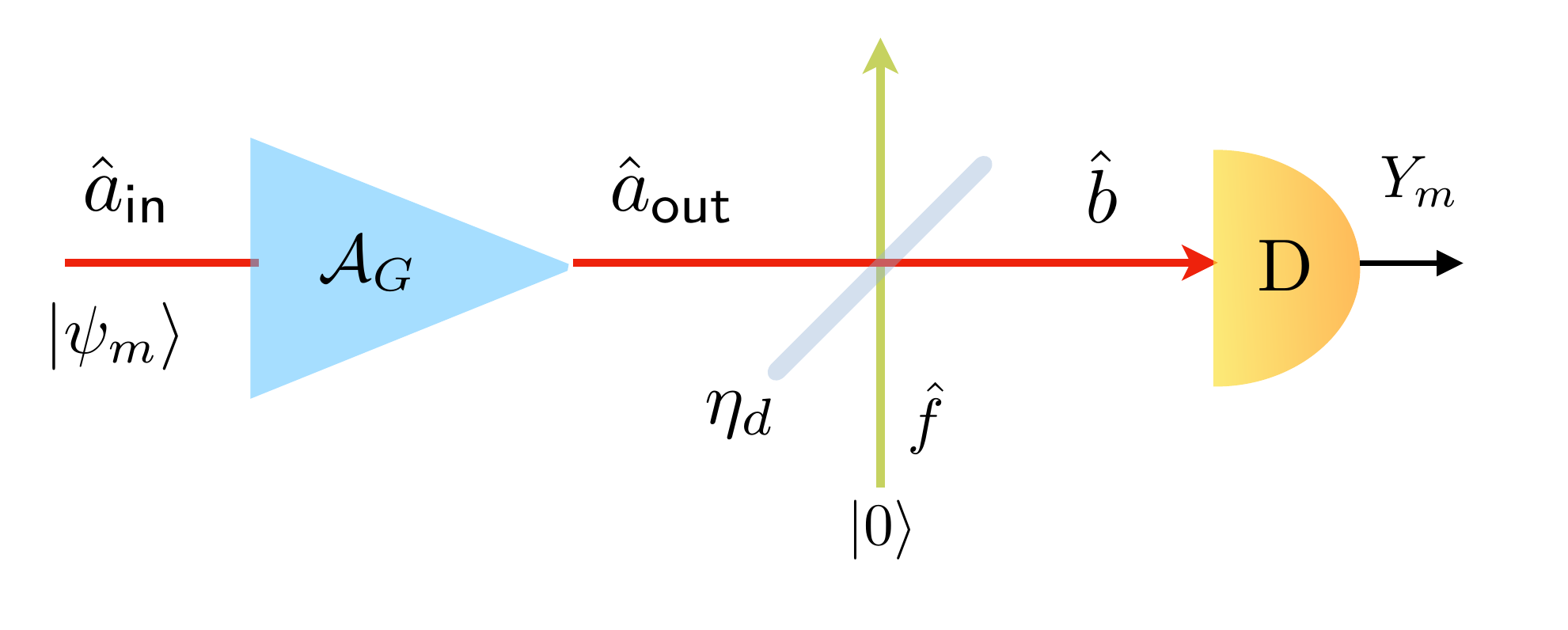}
  \caption{Gain estimation under inefficient detection: Each mode of a product signal-only probe $\otimes_{m=1}^M\ket{\psi_m}$  passes through a QLA $\mathcal{A}_G$. Detection with quantum efficiency $\eta_d$ is modeled by a beam splitter with mode $\hat{f}$ in vacuum and output mode $\hat{b}$ that is measured using an ideal photodetector D, resulting in photon count $Y_m$.}
  \label{fig:ineffpd}
\end{figure}

\emph{Practical quantum advantage---} To examine whether a quantum advantage can be demonstrated in the laboratory, we study the  estimation of $G$ using single-photon probes and photodetectors of efficiency $\eta_d <1$.  (See Fig.~\ref{fig:ineffpd}). For  any multimode number-state probe $\otimes_{m=1}^M\ket{n_m}$,  photon counting in each output mode remains the QFI-achieving measurement and the QFI can be obtained numerically  (Appendix~D.II). We also calculate the QFI of a coherent-state probe $\otimes_{m=1}^M \ket{\sqrt{N_m}}$ of the same $N$ and $M$ (Appendix D.I), and also the MSE of the unbiased estimator 
\begin{align} \label{Gchecklossy}
\check{G} &= \pars{\eta_d^{-1}Y +M}/\pars{N+M}
\end{align}generalizing that of Eq.~\eqref{Gcheck} (Appendix D.III).

\begin{figure*}[t] 
    \begin{subfigure}[b]{0.345\textwidth}\label{fig:4a}
    \centering
    {\includegraphics[trim=8mm 62mm 23mm 67mm, clip=true, scale=0.34]{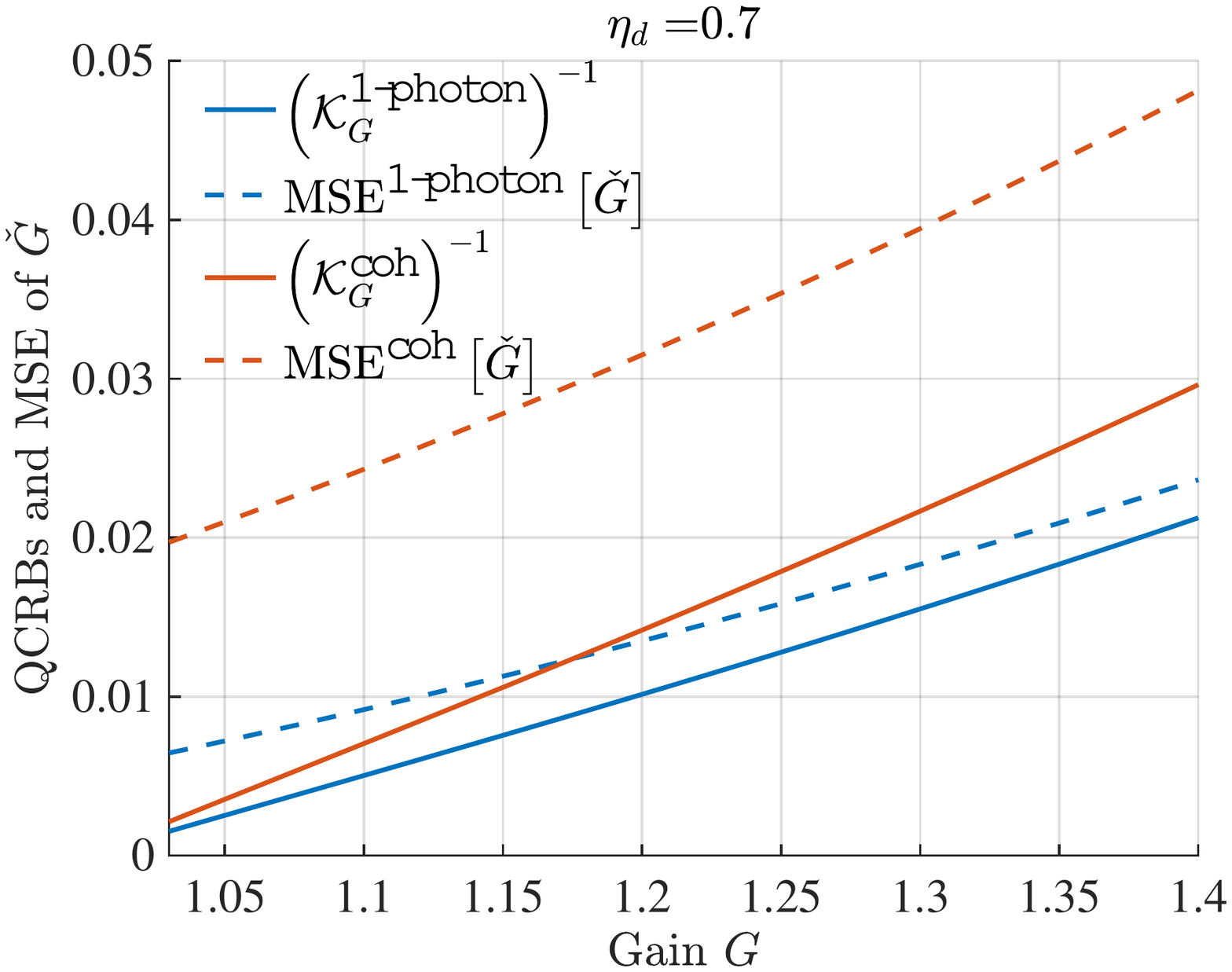}} 
    \end{subfigure}
     \begin{subfigure}[b]{0.322\textwidth}
    \centering
    {\includegraphics[trim=17.5mm 62mm 21mm 67mm, clip=true, scale=0.34]{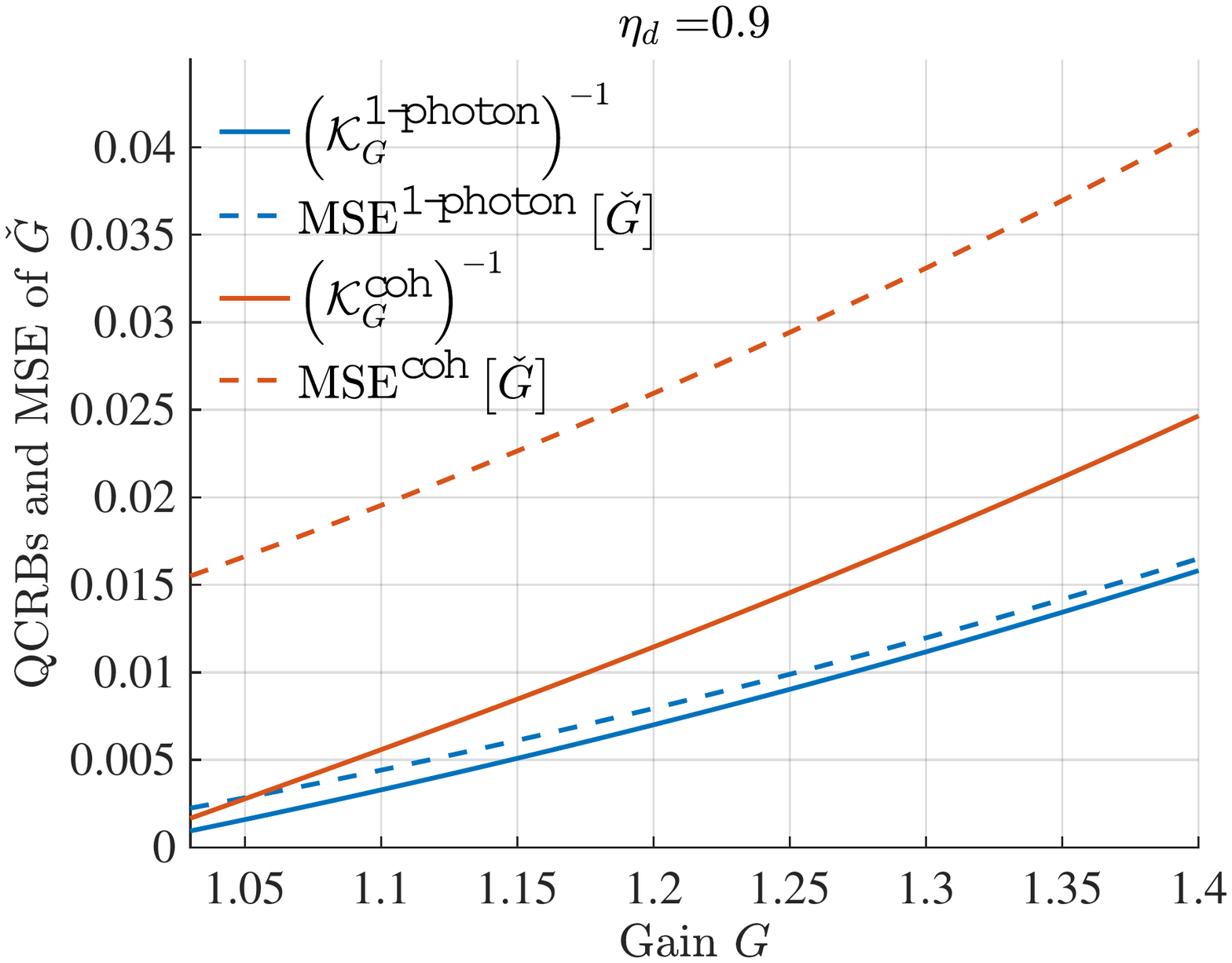}} 
    \end{subfigure}
     \begin{subfigure}[b]{0.322\textwidth}
    \centering
    {\includegraphics[trim=7mm 62mm 23mm 67mm, clip=true, scale=0.34]{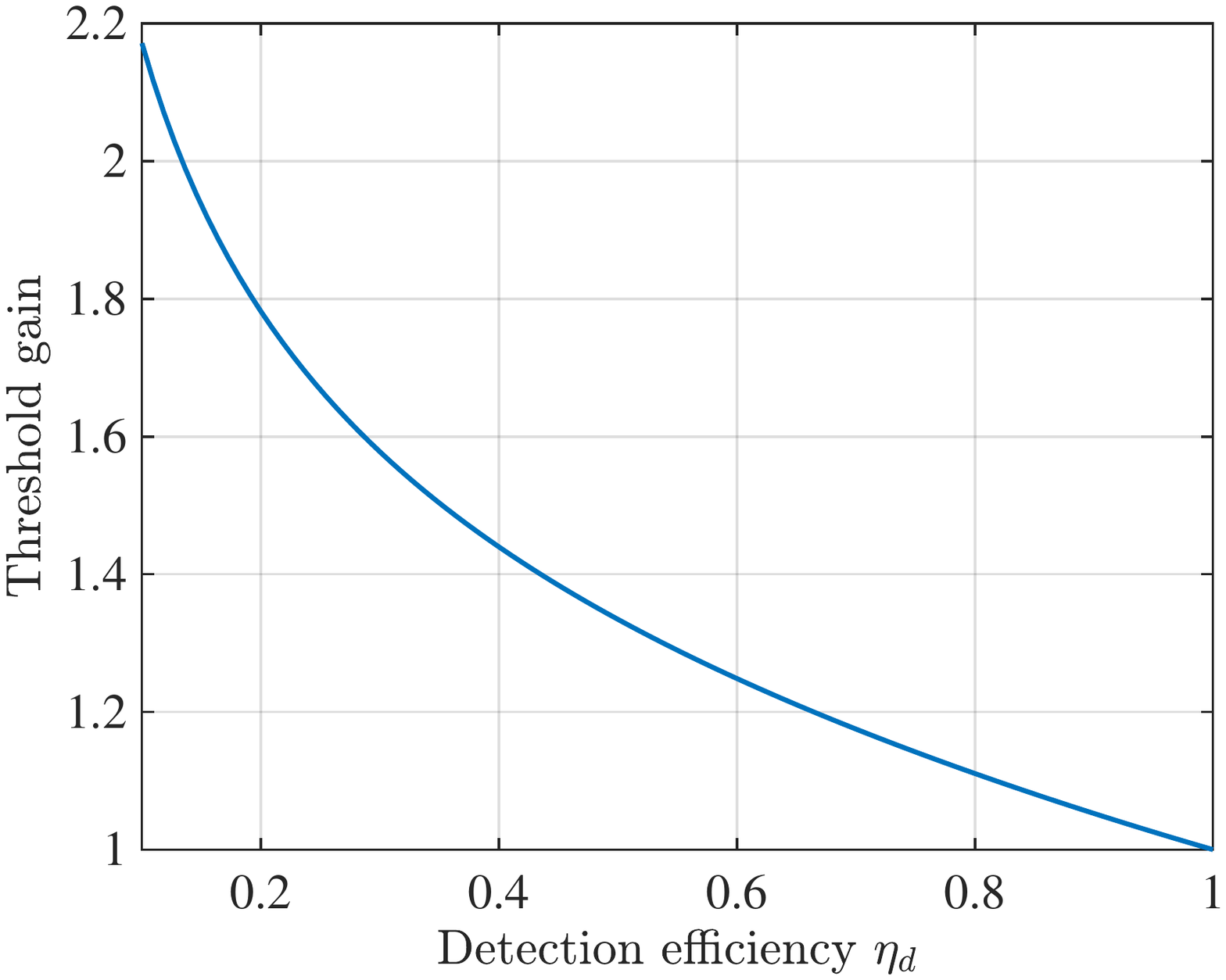}} 
    \end{subfigure}
    \caption{Performance of single-photon probes with inefficient detection: (Left) and (Center) -- QCRBs of multimode single-photon (blue solid) and coherent-state (red solid) probes along with the MSE of $\check{G}$ of Eq.\eqref{Gchecklossy} for single-photon (blue dashed) and coherent-state (red dashed) probes  for $\eta_d = 0.7$ (left) and $\eta_d = 0.9$ (center) with $M=N=20$.  (Right) -- The threshold gain beyond which single-photon probes and photon counting beat the coherent-state QCRB.}\label{fig:ineffdetplots}
\end{figure*}

Since single-photon states are  more readily prepared than multiphoton Fock states \cite{M-SSM20}, we compare their  performance relative to coherent states in Fig.~\ref{fig:ineffdetplots}. The MSE $\mathrm{MSE}^{\tsf{1-photon}}\bracs{\check{G}}$ of $\check{G}$ for single-photon probes (for which $M=N$) is always less than that for coherent states (Appendix~D.III), and Fig.~\ref{fig:ineffdetplots}, left and center).  Moreover, for each value of $\eta_d$, there is a threshold value of the gain (which is independent of $M$) beyond which $\mathrm{MSE}^{\tsf{1-photon}}\bracs{\check{G}}$ falls below the QCRB for coherent states (Fig.~\ref{fig:ineffdetplots}, right), so that a quantum advantage is guaranteed for sensing gain values known to lie beyond the threshold.

\emph{Energy-constrained Bures distance---}
As our final result, we derive the energy-constrained Bures distance \cite{Shi19uniform} between the amplifier channels $\cl{A}_G^{\otimes M}$ and $\cl{A}_{G'}^{\otimes M}$. This distance is one of several energy-constrained channel divergence measures between bosonic channels, with many applications in  quantum information and sensing \cite{PLO+17,PL17,Win17arxiv,Shi18ecdn,Shi19uniform,SWA+18, SH19,BD20,BDL+21,SSW20arxiv,TW16arxiv}. Its calculation is equivalent to minimizing the output fidelity $F\pars{\rho_G,\rho_{G'}}$ over all $M$-signal-mode probes \eqref{probe} with average signal energy $N$. We show (Appendix E) that this minimum equals
$F_{\rm min}\pars{\rho_G,\rho_{G'}} = \nu^M\bracs{\pars{1-\{N\}}\nu^{\lfloor N \rfloor} + \{N\}\nu^{\lfloor N \rfloor + 1}},$
where $\lfloor N \rfloor$ and $\{N\}$ are respectively the integer and fractional parts of $N$.
This results adds QLAs  to the short list of channels for which exact values of energy-constrained channel divergences are known and  also gives bounds on other divergences between QLAs \cite{Aud14}.

\emph{Discussion---}
We have delineated the optimal precision of sensing the gain of QLAs regardless of their implementation platform and  explicit physical realization. Our problem formulation constrained the average signal energy to equal $N$ but since the optimal QFI increases with $N$,  NDS states of average energy $N$ are optimal over all probes with average energy less than or equal to $N$. 

For multimode number-state probes, we identified a concrete quantum-optimal estimator and showed the in-principle feasibility of quantum-enhanced gain sensing using standard single-photon sources \cite{M-SSM20} and photon counting even under inefficient detection.  Additional loss in the signal path upstream of the QLA can also be accounted for by our calculation techniques. The use of brighter TMSV sources is expected to harness the photon contribution to the QFI of Eq.~\eqref{ndsqfi} even better, and finding good measurements and estimators for TMSV probes with imperfect detection is of great interest for future work. Our study can be generalized to the estimation of multiple \cite{LYL+19} and distributed \cite{GBB+20,*ZZ21} gain parameters.
The implications of our results for relativistic metrology problems \cite{AAF10,M-MFM11,*ABF14,*ABS+14} also remain to be explored.

Noisy attenuator channels (relevant to quantum illumination, noisy imaging, and quantum reading \cite{TEG+08,GMT+20,Pir11,OLP+21} among other applications), noisy amplifier channels (which model laser amplifiers with incomplete inversion \cite{Aga12quantum,Ser17qcv}), and additive noise channels (relevant to noisy continuous-variable teleportation \cite{BK98,*FSB+98}) are compositions of pure-loss channels with QLA channels. Our work here, together with complementary results in loss sensing \cite{Nai18loss},  is expected to be basic to a general theory of fundamental limits for sensing such noisy phase-covariant Gaussian channels, while highlighting the role of $M$ as an important resource therein.

\begin{acknowledgments}
This work is supported by the Singapore Ministry of Education Tier 1 Grants RG162/19 (S) and RG146/20, the NRF-ANR joint program (NRF2017-NRF-ANR004 VanQuTe), the National Research Foundation (NRF) Singapore under its NRFF Fellow program (Award No. NRF-NRFF2016-02), and  the FQXi R-710-000-146-720 Grant ``Are quantum agents more energetically efficient at making predictions?'' from the Foundational Questions Institute and Fetzer Franklin Fund (a donor-advised fund of Silicon Valley Community Foundation). Any opinions, findings and conclusions or recommendations expressed in this material are those of the author(s) and do not reflect the views of National Research Foundation or the Ministry of Education, Singapore.
\end{acknowledgments}

\onecolumngrid
\section*{APPENDICES}
\section{Appendix A: Action of the paramp unitary}

\noindent In this appendix, we compute the action of the paramp unitary $\hat{U}(\tau) := \exp\pars {-\iota t \hat{H}_I /\hbar} = \exp \bracs{\tau \pars{\hat{a}\hat{e} - \hat{a}^\dag \hat{e}^\dag}}$ on the states $\kets{n}\kete{0}$, which will enable us to compute $\Psi_\tau$ for arbitrary probes $\ket{\psi}_{AS}$. Recognizing that $\hat{U}(\tau)$ is a two-mode squeezing operation, we rewrite it in the alternative form (Cf. Eq.~(3.7.50) of \cite{BR02methods})
\begin{equation}
\begin{aligned} \label{disentU}
\hat{U}(\tau) =& \exp\bracs{\pars{\tanh \tau} \hat{a}^\dag \hat{e}^\dag} \, \exp\bracs{\pars{\ln \sech \tau}\pars{\hat{a}^\dag \hat{a} + \hat{e}\hat{e}^\dag}} \,\exp\bracs{-\pars{\tanh \tau} \hat{a} \hat{e}}.
\end{aligned}
\end{equation}
Upon expanding the exponentials appearing above into power series, we find that $\hat{U}(\tau)$ acts on the product state $\kets{n}\kete{0}$ as 
\begin{align} \label{actiononn0}
\hat{U}(\tau) \kets{n}\kete{0}&= \sech^{(n+1)} \tau\sum_{a=0}^\infty \sqrt{{n +a \choose a}} \tanh^a \tau \kets{n+a}\kete{a}, 
\end{align}
which appears above Eq.~(4) of the main text.

We recall the definition of a negative binomial distribution. Consider independent and identically distributed trials of flipping a coin with probability $p$ of landing heads. For integer $r \geq 1$, let  $X$ be the number of tails obtained before obtaining $r$ heads. The probability distribution of $X$ is
\begin{align}
P_X(x) = {x+r-1 \choose x} (1-p)^x\,p^r; \;\;x=0,1,\ldots
\end{align}
This distribution is called a negative binomial distribution and denoted $\mr{NB}(r,p)$ \cite{RS15probstats}. Comparing with Eq.~\eqref{actiononn0}, we see that the number $a$ of photons  added by the amplifier to the $S$ and $E$ modes is distributed according to the $\mr{NB}(n+1,\sech^2\tau)$ distribution when the input is $\kets{n}\kete{0}$.

\section{Appendix B: Fidelity between output states of $ASE$ and between the output states of $AS$ for NDS probes} \label{sec:fidelity}

\noindent In this appendix, we prove Eq.~(5) of the main text. We first need the following combinatorial lemma.
\begin{lemma} \label{lem:comb}
Given an $M$-vector $\mb{n} = \pars{n_1, \ldots, n_M}$  of nonnegative integers and an integer $a \geq 0$, we have
\begin{align} \label{lemma}
\sum_{\mb{a} \geq \mb{0}: \tr \mb{a} =a} \bracs{\prod_{m=1}^M {n_m + a_m \choose a_m}} &= {\tr \mb{n}+ M-1 + a \choose a},
\end{align}
where  $\mb{a} = \pars{a_1, \ldots, a_M} \geq \mb{0}$ is another $M$-vector, and the trace of a vector $\mb{x}$ is defined to be $\tr \mb{x} := \sum_{m=1}^M x_m$.
\begin{proof} 
For  integer $n\geq 0$, we have the Taylor series expansion (sometimes called the ``negative binomial theorem''):
\begin{align} \label{taylor}
(1-x)^{-n} = \sum_{k=0}^\infty {n +k - 1 \choose k}\,x^k,
\end{align}
which is valid for $\abs{x} <1$. For any  $\mb{n} \geq \mb{0}$, we can write
\begin{align}
\prod_{m=1}^M (1-x)^{-(n_m +1)} = (1-x)^{- \pars{\tr \mb{n} + M}}.
\end{align}
For integer $a \geq 0$, consider the coefficient of $x^a$ on each side of this identity. Applying \eqref{taylor} to each of the factors on the left-hand side and separately to the right-hand side, we see that this coefficient equals in turn the expressions on either side of Eq.~\eqref{lemma}. 
\end{proof}
\end{lemma}

The overlap $\braket{{\Psi_\tau}}{{\Psi_{\tau'}}}$ between output states of the $ASE$ system equals 
\begin{align}
\braket{{\Psi_\tau}}{{\Psi_{\tau'}}} &= \sum_{\mb{a}\geq\mb{0}} \varbraket{\psi_{\tau;\mb{a}}}{\psi_{\tau';\mb{a}}} \label{app:fidelity}\\
&= \sum_{\mb{a} \geq \mb{0}} \sum_{\mb{n} \geq \mb{0}} p_{\mb{n}} \, \sqrt{A_\tau (\mb{n},\mb{a} ) A_{\tau'}(\mb{n},\mb{a})} \\
&= \sum_{\mb{n} \geq \mb{0}} p_{\mb{n}} \pars{\sech \tau \sech \tau'}^{n+M} \sum_{\mb{a} \geq \mb{0}} \bracs{\prod_{m=1}^M {n_m + a_m \choose a_m} \pars{\tanh \tau \, \tanh \tau'}^{a_m}},
\end{align}
where, as before, $n := \tr \mb{n}$. We now rearrange the sum over $\mb{a}$ in terms of $a :=  \tr \mb{a}\geq 0$ and use the lemma to get
\begin{align}
&\braket{{\Psi_\tau}}{{\Psi_{\tau'}}} \\
&= \sum_{\mb{n} \geq \mb{0}} p_{\mb{n}} \pars{\sech \tau \sech \tau'}^{n+M} \sum_{a=0}^\infty \sum_{a \geq \mb{0}: \tr \mb{a} = a}  \bracs{\prod_{m=1}^M {n_m + a_m \choose a_m} } \pars{\tanh \tau \, \tanh \tau'}^a\\
&= \sum_{\mb{n} \geq \mb{0}} p_{\mb{n}} \pars{\sech \tau \sech \tau'}^{n+M} \sum_{a=0}^\infty {n+M -1 +a \choose a}\pars{\tanh \tau \, \tanh \tau'}^a \\
&= \sum_{\mb{n} \geq \mb{0}} p_{\mb{n}} \pars{\cosh \tau \,\cosh \tau' - \sinh \tau \, \sinh \tau'}^{-(n+M)} \\
&= \sum_{n=0}^\infty p_n \bracs{\sech(\tau'-\tau)}^{n+M} \\
&\equiv \sum_{n=0}^\infty p_n \nu^{n+M}, \label{app:ndsfidelity}
\end{align}
which establishes Eq.~(5) of the main text. As argued there, the right-hand side of Eq.~\eqref{app:fidelity} is also the fidelity $F\pars{\rho_\tau, \rho_{\tau'}}$ between the output states of the $AS$ system alone when the probe $\ket{\psi}_{AS}$ is NDS. 

\section{Appendix C: Fisher Information of the Schmidt-bases measurement for NDS probes}

\noindent Suppose that an NDS probe $\ket{\psi} = \sum_{\mb{n} \geq \mb{0}} \sqrt{p}_{\mb{n}}\ket{\chi_{\mb n}}_A \ket{\mb n}_S$ is used and that we measure at the output of the amplifier its Schmidt bases, i.e., the basis $\left\{\keta{\chi_{\mb n}}\right\}$ in the ancilla system and also the photon number in each of the $M$ output signal modes. The measurement result is denoted $(\mb{X},\mb{Y})$, where $\mb{X}=(X_1, \ldots, X_M)$ is the index of the measurement result on $A$ ($\mb{X} = \mb{x}$ if the result $\keta{\chi_{\mb{x}}}$ is obtained) and $\mb{Y} = (Y_1,\ldots, Y_M)$ denotes the outcome of the photon number measurement in the $M$ signal modes. From Eq.~(4) of the main text, the joint distribution of this observation is 
\begin{align} \label{app:measpmf}
P(\mb{x}, \mb{y}; \tau) &=  p_{\mb{x}} A(\mb{x}, \mb{y} - \mb{x}) =  p_{\mb{x}} \pars{\prod_{m=1}^M {y_m \choose x_m} \sech^{2(x_m+1)}\tau \tanh ^{2 (y_m - x_m)}\tau}.
\end{align}
Using the general expression 
\begin{align} \label{FI}
\cl{J}_{\theta} \bracs{Z} = - \mathbb{E} \bracs{{\partial_{\theta}^2 \ln P(Z;\theta)}}
\end{align}
 for the Fisher information on a parameter $\theta$ of a measurement $Z$ \cite{Kay93ssp1}, we have 
\begin{align}
\cl{J}_\tau \bracs{\mb{X},\mb{Y}} &= 2 \sum_{m=1}^M \bracs{\pars{\mathbb{E}[X_m]+1} \sech^2\tau + {\mathbb{E}[Y_m -X_m]}\pars{\sech^2 \tau + \csch^2 \tau}} \\
&= 4 \sum_{m=1}^M \bracs{ \mathbb{E}[X_m] + 1} = 4(N+M). \label{countingFI}
\end{align}
Here, we have used the fact that, conditioned on $X_m =x_m$, $Y_m \thicksim \mr{NB}(x_m+1, \sech^2 \tau)$ so that $\mathbb{E}[Y_m -X_m] = \pars{\mathbb{E}[X_m] +1}\sinh^2 \tau$, and the energy constraint  is used in the last step. Since Eq.~\eqref{countingFI} coincides with the NDS-probe QFI of Eq.~(6) of the main text, we see that this measurement is quantum-optimal for any such probe. 

\section{Appendix D: Effect of nonunity detection efficiency}

\noindent Consider the measurement configuration depicted in Fig.~3 of the main text. The annihilation operator of the measured mode equals
\begin{equation}
\begin{aligned} \label{app:detmode}
\hat{b} &= \sqrt{\eta_d}\, \hat{a}_{\tsf{out}} + \sqrt{1-\eta_d}\, \hat{f} \\
&= \sqrt{\eta_d G} \,\hat{a}_{\tsf{in}} + \sqrt{\eta_d(G-1)} \,\hat{e}_{\tsf{in}}^\dag + \sqrt{1-\eta_d} \,\hat{f}.
\end{aligned}
\end{equation}
The photon number operator of $\hat{b}$ becomes
\begin{align} \label{app:numbop}
\hat{N}_b &= \eta_d\,\hat{N}_{\tsf{out}} + \sqrt{\eta_d (1-\eta_d)}\pars{\hat{a}_{\tsf{out}}^\dag \hat{f} + \hat{a}_{\tsf{out}} \hat{f}^\dag} + (1-\eta_d) \hat{f}^\dag\hat{f},
\end{align}
where $\hat{N}_{\tsf{out}} = \hat{a}_{\tsf{out}}^\dag\hat{a}_{\tsf{out}}$.

\subsection{I.~Classical baseline}

\noindent Suppose the coherent-state probe $\kets{\sqrt{N}}$ is used in Fig.~3 of the main text, but arbitrary measurements are allowed on the mode $\hat{b}$ downstream of the beam splitter. This corresponds to the case where a system loss (including detection efficiency) $\eta_d$ is present in the system. The smallest MSE (optimized over all quantum measurements) in estimating $G$ using classical probes is set by the QCRB of the state in the mode $\hat{b}$. Using Eq.~\eqref{app:detmode}, and noting that the modes $\hat{e}_{\tsf{in}}$ and $\hat{f}$ are in the vacuum state, we find that the mean vector and Wigner covariance matrix of the quadratures $\hat{q} = \pars{\hat{b} + \hat{b}^\dag}/\sqrt{2}$ and $\hat{p} = \pars{\hat{b} - \hat{b}^\dag}/\sqrt{2}i$ of the $\hat{b}$ mode are given respectively by
\begin{equation}\label{app:csmeancov}
\begin{aligned}
\boldsymbol{\mu}_G &= \pars{\sqrt{2\eta_d G N}, 0}^{\trans}, \\
\boldsymbol{\Sigma}_G &= \bracs{\eta_d(G-1) + 1/2} \mathbb{1}_2,
\end{aligned}
\end{equation}
where $\mathbb{1}_2$ is the $2 \times 2$ identity matrix. Since $\hat{b}$ is in a Gaussian state (in fact, a displaced thermal state), we can invoke the results of Ref.~\cite{PJT+13} on computing the QFI of parameters of Gaussian states. Using Eqs.~\eqref{app:csmeancov}, we obtain after some computation the QFI
\begin{align} \label{app:csqfiineffdet}
\cl{K}_G^{\tsf{coh}} &= \frac{\eta_d N}{G\bracs{2\eta_d (G-1) + 1}} + \frac{\eta_d}{(G-1)\bracs{\eta_d(G-1)+1}}
\end{align}
on $G$ in the lossy regime, which is displayed in Fig.~4 (left and center plots) of the main text. As in the lossless case, it is unknown if  this QFI can be achieved on a finite sample using a $G$-independent measurement.

\subsection{II.~Quantum Fisher information for number-state probes}

\noindent Consider the scenario depicted in Fig.~3 of the main text for the number-state probe $\ket{\psi}_S = \otimes_{m=1}^M \ket{n_m}_{S_m}$.  The ultimate limit on the mean squared error of estimating $G$ is set by the QCRB of the state family in the set of modes $\left\{\hat{b}_m\right\}_{m=1}^M$. Since the  state of those modes is no longer Gaussian, we cannot appeal to the literature on parameter estimation for Gaussian-state families. From the action of the QLA on number-state inputs written above Eq.~(4) of the main text, however, the output state of the amplifier when $\ket{N}_S$ is input can be written as:
\begin{align} \label{app:rhotau}
\rho_\tau &=  \sech^{2(N+1)} \tau \sum_{a=0}^\infty {N +a \choose a} \tanh^{2a} \tau \kets{N+a}\bra{N+a}.
\end{align}
Since the beam splitter acts on number states according to $\kets{n}\bra{n} \mapsto \sum_{k=0}^n {n \choose k} \eta_d^k (1 -\eta_d)^{n-k} \kets{k}\bra{k}$, the state of the $\hat{b}$ mode becomes
\begin{equation}\label{app:rhotau'}
\begin{aligned} 
\rho_{\tau}'&= \sum_{k=0}^\infty \pars{\sech^{2(N+1)} \tau \sum_{a=\mr{max}(k-N,0)}^\infty {N+a \choose a} {N+a \choose k} \eta_d^k (1-\eta_d)^{N+a-k} \tanh^{2a} \tau} \kets{k}\bra{k} \\
&\equiv \sum_{k=0}^\infty P_{\tau}(k) \kets{k}\bra{k}.
\end{aligned}
\end{equation}
Since the state family $\left\{\rho_{\tau}'\right\}$ is diagonal in the number basis, the QFI on $\tau$ equals the FI of the family of photon number distributions $\left\{P_\tau(k); k = 0,1,2,\ldots\right\}$ and is achieved by photodetection. Using Eq.~\eqref{FI}, this quantity is
\begin{align}
\cl{K}_\tau &= \cl{J}_\tau[Y] \\
&= -\sum_{k=0}^\infty P_\tau(k) \partial^2_{\tau} \bracs{\ln P_\tau(k)} \\
&= \sum_{k=0}^\infty \bracs{ P_\tau^{-1}(k) \bracs{\partial_{\tau} P_\tau (k) }^2 - \partial_\tau^2 P_\tau(k)}.
\end{align}
Evaluating the derivatives results in the somewhat unwieldy expressions
\begin{align}
\partial_{\tau} P_\tau (k) = 2\,\eta_d^k \sech^{2(N+2)} \tau &\sum_{a=\mr{max}(k-N,0)}^\infty {N +a \choose a} {N +a \choose k} 
(1-\eta_d)^{N+a-k}\bracs{a - (N+1) \sinh^2 \tau} \tanh^{2a-1}\tau, \\
\partial^2_{\tau} P_\tau (k) = 2\,\eta_d^k \sech^{2(N+3)} \tau &\sum_{a=\mr{max}(k-N,0)}^\infty {N +a \choose a} {N +a \choose k} (1-\eta_d)^{N+a-k}\;\nonumber\\
\times &  \bracs{2 (N+1)^2 \sinh^4 \tau - \pars{4aN + N + 6a+1}\sinh^2 \tau + 2a^2 - a} \tanh^{2a-1}\tau,
\end{align}
using which the QFI $\cl{K}_\tau$ can be evaluated numerically. The QFI $\cl{K}_G$ on $G$ follows as $\cl{K}_G = \cl{K}_\tau/ [4G(G-1)].$ The QFI for a multimode number-state probe is  the sum of terms of the above form for each mode. The result for multimode single-photon probes is displayed in Fig.~4 (left and center plots) of the main text.

\subsection{III.~Mean squared error of the estimator $\check{G}$ of Eq.~(9) of the main text}

\noindent In this section, we  compute the MSE achieved by the estimator 
\begin{align} \label{app:Gcheck}
\check{G} = \frac{\pars{\sum_{m=1}^M Y_m}/{\eta_d} + M}{N+M}
\end{align}
on $M$-mode number-state and coherent-state probes with average total energy $N$, where the $\left\{Y_m\right\}$ denote the observed photocounts in the modes $\left\{\hat{b}_m\right\}$. Using Eq.~\eqref{app:numbop} and the fact that the modes $\hat{e}$ and $\hat{f}$ are in the vacuum state, we have after some algebra that the first and second moments of the photocount in each mode satisfy (we omit mode subscripts to reduce clutter):
\begin{align} 
\mean{Y} &= \mean{\hat{N}_{b}} = \eta_d \mean{\hat{N}_{\tsf{out}}}, \label{app:meanY} \\
\mean{Y^2} &= \mean{\hat{N}_b^2} =  \eta_d^2 \mean{\hat{N}^2_{\tsf{out}}} + \eta_d(1-\eta_d)\mean{\hat{N}_{\tsf{out}}}, \label{app:2ndmomY}
\end{align}
where $\hat{N}_{\tsf{out}} = \hat{a}^\dag_\tsf{out} \hat{a}_{\tsf{out}}$.
The mean and second moment of $\hat{N}_{\tsf{out}}$ are obtained from the Heisenberg-evolution equations for a QLA given in the main text:
\begin{align}
\mean{\hat{N}_{\tsf{out}}}  &= G\mean{\hat{N}_{\tsf{in}}} + G-1, \label{app:meanNout}\\
\mean{\hat{N}_{\tsf{out}}^2}  &= G^2\mean{\hat{N}^2_{\tsf{in}}} + 3G(G-1)\mean{\hat{N}_{\tsf{in}}} + (G-1)(2G-1), \label{app:2ndmomNout}
\end{align}
where $\hat{N}_{\tsf{in}} = \hat{a}_{\tsf{in}}^\dag \hat{a}_{\tsf{in}}$. Using Eqs.~\eqref{app:meanY} and \eqref{app:meanNout}, we see that the estimator of Eq.~\eqref{app:Gcheck} is unbiased  for any probe state. 

If the probe is of product form, the variance of $\check{G}$ can be obtained by applying Eqs.~\eqref{app:2ndmomY} and \eqref{app:2ndmomNout}. In particular, for a number-state probe $\otimes_{m=1}^M \ket{n_m}$ with $N = \sum_{m=1}^M n_m$, we find that
\begin{align} \label{MSEGchecknum}
\mr{MSE}^{\tsf{num}}\bracs{\check{G}} = \frac{G(G-1)}{N+M} + \frac{1-\eta_d}{\eta_d(N+M)}\bracs{G - \frac{M}{N+M}},
\end{align}
where the first term corresponds to the QCRB under ideal photodetection and the second term is the excess error introduced due to the inefficiency.

For a coherent-state probe $\otimes_{m=1}^M \ket{\sqrt{N_m}}$ with $N = \sum_{m=1}^M N_m$, we have
\begin{align} \label{MSEGcheckcoh}
\mr{MSE}^{\tsf{coh}}\bracs{\check{G}} = \frac{G(G-1)}{N+M} + \frac{G^2 N}{(N+M)^2} +\frac{1-\eta_d}{\eta_d(N+M)}\bracs{G - \frac{M}{N+M}}.
\end{align}
Again, the first term corresponds to the quantum-optimal error while the middle term represents an additional error owing to the suboptimality of the coherent-state probe. In fact, the sum of the first two terms is strictly greater than the QCRB ${\cl{K}_G^{\tsf{coh}}}^{-1}$ (cf. Eq.~(7) of the main text), corresponding to the fact that photon counting is not a QFI-achieving measurement for coherent-state probes. Finally the last term is the excess error owing to inefficient detection, which is identical to the corresponding term in Eq.~\eqref{MSEGchecknum}.  These results on the MSE of $\check{G}$ are plotted in Fig.~4 of the main text (left and center plots) for coherent states and multimode single-photon probes, for which $M=N$. For each  $\eta_d$, Eq.~\eqref{MSEGchecknum} drops below the coherent-state QCRB (inverse of Eq.~\eqref{app:csqfiineffdet}) whenever $G$ is greater than a threshold value -- this value is plotted in Fig.~4 (right plot) of the main text.

\section{Appendix E: Energy-constrained Bures distance between amplifier channels}
In this appendix, we obtain the energy-constrained Bures distance between two given product amplifier channels $\cl{A}_G^{\otimes M}$ and $\cl{A}_{G'}^{\otimes M}$ acting on $M$ modes. To define this quantity, consider the ancilla-assisted channel discrimination problem  in which each signal mode of a probe of $M$ of signal modes and total energy $N$ entangled with an  arbitrary ancilla $A$ queries a  black box containing one of the quantum-limited amplifiers $\cl{A}_G$ or $\cl{A}_{G'}$.  Since the total signal energy is constrained,  we can ask what probe state maximizes a chosen state distinguishability measure at the output.  Several such channel distinguishability measures under an energy constraint, known as energy-constrained channel divergences, have been proposed recently, e.g.,  the energy-constrained diamond distance and  the \emph{energy-constrained Bures  (ECB) distance} among others, with several applications in bosonic quantum information \cite{PLO+17,PL17,Win17arxiv,Shi18ecdn,Shi19uniform,SH19,SSW20arxiv,BDL+21,Shi19uniform,SWA+18,TW16arxiv}. We focus here on the ECB distance \cite{Shi19uniform}, defined for the problem at hand as follows: The ECB distance between $\cl{A}_G^{\otimes M}$ and $\cl{A}_{G'}^{\otimes M}$ is given by the expression:
\begin{align} \label{ecbdistance}
B_N\pars{\cl{A}_\tau^{\otimes M},\cl{A}_{\tau'}^{\otimes M}} := 
\sup_{\rho_{AS} : \Tr \rho_{AS}\hat{I}_A\otimes  \hat{N}_S  =N} \sqrt{1 - F\pars{{\rm id}\otimes\cl{A}_\tau^{\otimes M}(\rho_{AS}),{\rm id}\otimes\cl{A}_{\tau'}^{\otimes M}(\rho_{AS})}},
\end{align}
where $F$ is the fidelity, $A$ is an arbitrary ancilla system, id is the identity channel on $A$, $\hat{N}_S$ is the total photon number operator on $S$, and the optimization is over all states $\rho_{AS}$ of $AS$ with signal energy $N$. The formulation of Eq.~\eqref{ecbdistance} differs slightly from that in \cite{Shi19uniform} in using an equality energy constraint, and is normalized to lie between 0 and 1 rather than 0 and $\sqrt{2}$. As we show, the two definitions give the same ECB distance up to normalization for the problem at hand.

First, note that an arbitrary $\rho_{AS}$ satisfying the constraint can be purified using an additional ancilla to give a probe of the form of Eq.~(1) of the main text, so that the optimization over probe states of $AS$ can be restricted to pure states. Maximizing the output Bures distance \label{ecbdist} is equivalent to minimizing the output fidelity. We claim that an optimal probe $\ket{\psi}_{AS}$ must be of the NDS form, i.e., the $\{\ket{\chi_{\mb n}}_A\}$ appearing in Eq.~(1) of the main text must be  orthonormal.  To see this, recall from Sec.~\ref{sec:fidelity} that if the environment modes are accessible, the output fidelity of the purified states takes the value given in Eq.~\eqref{app:ndsfidelity}. Since the fidelity between the accessible outputs $\rho_{\tau}$ and $\rho_{\tau'}$ on the $AS$ system cannot be less than between their purifications, we have
\begin{align} \label{fidbound}
F\pars{\rho_{\tau},\rho_{\tau'}} \geq \nu^M \sum_{n=0}^\infty p_n \nu^{n}.
\end{align}
On the other hand, we argued in the main text that the right-hand side is achieved by any NDS probe with the same photon number distribution $\left\{p_n\right\}$, so the optimum fidelity is achieved on an NDS probe. Alternatively, this conclusion  follows directly from the argument of Section 12 of \cite{SWA+18} that NDS probes are optimal for discriminating between any pair of phase-covariant channels, which amplifier channels are.

Since $M$ is fixed, we need to minimize  $\sum_n p_n \nu^n$ under the energy constraint $\sum_n n p_n  =N$. Since the function $x \mapsto \nu^x$ is convex, we have $\sum_n p_n \nu^n \geq \nu^{\sum_n n p_n} = \nu^N$ for any probe. If $N$ is an integer, this value is achieved iff the probe satisfies $p_N =1$, e.g., the probe could be a multimode number state of total photon number $N$. For general $N$, we can reprise an argument from \cite{Nai18loss} as follows:
For any $\{ p_n\}$ satisfying the energy constraint, let ${A}_{\downarrow} = \sum_{n \leq \lfloor N \rfloor} p_n$, and $A_{\uparrow} = 1 - {A}_{\downarrow}$. For ${N}_{\downarrow} = A_{\downarrow}^{-1} \sum_{n \leq \lfloor N \rfloor} n\,p_n \leq \lfloor N \rfloor$ and 
${N}_{\uparrow} = A_{\uparrow}^{-1}\sum_{n \geq \lceil N \rceil} n\,p_n \geq \lceil N \rceil$, we have   $A_\downarrow\,N_\downarrow + A_\uparrow\,N_\uparrow = N$. Convexity of $x \mapsto \nu^x$ implies that
$\sum_n p_n \nu^n \geq {A}_{\downarrow} \,\nu^{N_{\downarrow}} + {A}_{\uparrow} \,\nu^{N_{\uparrow}}$.
Convexity  also implies that the chord joining $({N}_{\downarrow}, \nu^{N_{\downarrow}})$ and $({N}_{\uparrow}, \nu^{N_\uparrow})$
lies  above that joining $(\lfloor N \rfloor, \nu^{\lfloor N \rfloor})$ and $(\lceil N \rceil , \nu^{\lceil N \rceil})$ in the interval $\lfloor N \rfloor \leq x \leq \lceil N \rceil$ -- this follows from the fact that the intersection of the epigraph of $x \mapsto \nu^x$ and the region above the line joining $(\lfloor N \rfloor, \nu^{\lfloor N \rfloor})$ and $(\lceil N \rceil , \nu^{\lceil N \rceil})$ is the intersection of two convex sets and hence is also convex.  Denote the fractional part of $N$ by $\left\{N\right\} = N - \lfloor N\rfloor$. Since the energy constraint  can  be satisfied by taking ${N}_\downarrow = \lfloor N \rfloor, {N}_\uparrow = \lceil N \rceil, p_{\lfloor N \rfloor} = 1 -\{N\}$, and $p_{\lceil N \rceil} = \{N\}$, the energy-constrained minimum fidelity equals:
\begin{align} \label{ecminf}
F^{\rm min}\pars{\rho_{\tau},\rho_{\tau'}}  = \nu^M \bracs{\pars{1- \{N\}}\,\nu^{\lfloor N \rfloor} +  \{N\}\,\nu^{\lceil N \rceil}}.
\end{align}
The ECB distance $B_N\pars{\cl{A}_\tau^{\otimes M},\cl{A}_{\tau'}^{\otimes M}} $ then follows from Eq.~\eqref{ecbdistance}. Since the  ECB distance is an increasing function of $N$, it equals (up to normalization) the ECB distance defined using an inequality constraint in \cite{Shi19uniform}. 

For quantifying the advantage of using quantum probes, we can define a \emph{classical} energy-constrained Bures distance (CECB distance) analogous to \eqref{ecbdistance} except that the probes $\rho_{AS}$ are restricted to be in the set  of classical states of the form
\begin{align} \label{classical probe}
\rho_{AS} = \int_{\mathbb{C}^{M'}} \diff^{2M'} \balpha \int_{\mathbb{C}^{M}} \diff^{2M} \bbeta \,P(\balpha,\bbeta) \ket{\balpha}\bra{\balpha}_A\otimes\ket{\bbeta} \bra{\bbeta}_S,
\end{align}
where $\bbeta = \pars{\alpha_S^{(1)}, \ldots, \alpha_S^{(M)}} \in \mathbb{C}^M$ indexes $M$-mode coherent states $\kets{\bbeta}$ of $S$, $\balpha = \pars{\alpha_A^{(1)}, \ldots, \alpha_A^{(M')}}\in \mathbb{C}^{M'}$ indexes $M'$-mode coherent states $\kets{\balpha}$ of $A$, and $P\pars{\balpha,\bbeta} \geqslant 0$ is a probability distribution. 
Additionally, the signal energy constraint implies that $P\pars{\balpha,\bbeta}$ should satisfy
\begin{align} \label{CEC}
\int_{\mathbb{C}^{M'}} \diff^{2M'} \balpha \int_{\mathbb{C}^M} \diff^{2M} \bbeta\, P(\balpha,\bbeta) \pars{\sum_{m=1}^M \abs{\alpha_S^{(m)}}^2}  = {N}.
\end{align}
 Denote by $\cl{S}^{\tsf{cl}}_N$ the set of classical states satisfying the energy constraint \eqref{CEC}. The CECB distance is then defined as
\begin{equation}
\begin{aligned} \label{cecbdistance}
B_N^{\tsf{cl}}\pars{\cl{A}_\tau^{\otimes M},\cl{A}_{\tau'}^{\otimes M}} := 
 \sup_{\rho_{AS} \in \cl{S}_N^{\tsf{cl}}} \sqrt{1 - F\pars{{\rm id}\otimes\cl{A}_\tau^{\otimes M}(\rho_{AS}),{\rm id}\otimes\cl{A}_{\tau'}^{\otimes M}(\rho_{AS})}}.
\end{aligned}
\end{equation}
Consider first a coherent-state probe $\kets{\sqrt{N}}$ of a single signal mode. Since the  corresponding output state $\rho_\tau = \Tr_E \hat{U}(\tau)\, \ket{\sqrt{N}}\bra{\sqrt{N}}_S \otimes\ket{0}\bra{0}_E\, \hat{U}^\dag(\tau)$ is a displaced thermal state, we can use results on the fidelity between Gaussian states \cite{MM12} to compute the fidelity $F\pars{\rho_\tau, \rho_{\tau'}}$:
\begin{align} \label{CSfidelity}
F\pars{\rho_\tau, \rho_{\tau'}} &= \sech\pars{\tau' - \tau}\, \exp\bracs{ \minus \frac{\pars{\cosh \tau' - \cosh \tau}^2}{2\pars{\sinh^2 \tau' + \sinh^2 \tau + 1}} {N}}
\end{align}
For a coherent-state $\ket{\sqrt{N_1}} \otimes \cdots \otimes \ket{\sqrt{N_M}} \in \cl{S}^{\tsf{cl}}_N$, it follows from Eq.~\eqref{CSfidelity} that the output fidelity is
\begin{align} \label{cecminf}
F^{\tsf{coh}} = \nu^M\, \exp\bracs{ - \frac{\pars{\cosh \tau' - \cosh \tau}^2}{2\pars{\sinh^2 \tau' + \sinh^2 \tau + 1}} {N}}.
\end{align}
The strong concavity of the fidelity \cite{NC00} over mixtures and the convexity with respect to $N$ of the exponential appearing in \eqref{cecminf} imply that the above expression is the minimum fidelity over all probes in $\cl{S}^{\tsf{cl}}_N$. The minimum output Bures distance over classical probes then follows from Eq.~\eqref{cecbdistance}.
Note that the dependence on $M$ of the minimum fidelity appears in both the quantum \eqref{ecminf} and classical \eqref{cecminf} expressions as the factor $\nu^M$.

\bibliography{../../RNmasterbib}



\end{document}